\documentclass[12pt]{article}
\usepackage{mathrsfs} \usepackage{amsmath} \usepackage{amssymb} \usepackage{slashed}
\usepackage{graphicx} \usepackage{caption2}
\begin{document}
\newcommand{\ba}{\begin{eqnarray}} \newcommand{\ea}{\end{eqnarray}}
\newcommand{\be}{\begin{equation}} \newcommand{\ee}{\end{equation}}
\renewcommand{\figurename}{Figure.} \renewcommand{\captionlabeldelim}{.~}
\renewcommand{\thefootnote}{\fnsymbol{footnote}}

\vspace*{1cm}
\begin{center}
 {\Large\textbf{A Model of the Matter-antimatter Asymmetry and Cold Dark Matter with $U(1)_{B-L}\otimes U(1)_{D}$}}

\vspace{1cm}
 \textbf{Wei-Min Yang}

\vspace{0.4cm}
 \emph{Department of Modern Physics, University of Science and Technology of China, Hefei 230026, P. R. China}

\vspace{0.2cm}
 \emph{E-mail: wmyang@ustc.edu.cn}
\end{center}

\vspace{1cm}
\noindent\textbf{Abstract}: I suggest an effective model between the GUT and the electroweak scale. It only introduces the two symmetries of $U(1)_{B-L}$ and $U(1)_{D}$ besides the SM groups. The two symmetries are individually broken at the reheating temperature of the universe of $10^{12}$ GeV and the scale of $3\sim 4$ TeV. The model can simultaneously accommodate the tiny neutrino masses, the matter-antimatter asymmetry and the cold dark matter (CDM). In particular, the model gives some interesting results and predictions, for instance, the neutrinos are Dirac nature and their masses are related to the $U(1)_{D}$ breaking, the size of the matter-antimatter asymmetry is closely related to the mass hierarchy of the quarks and charged leptons, the CDM mass is probably in the range of $250\sim 350$ GeV. Finally, it is feasible to test the model in future collider experiments.

\vspace{1cm}

\vspace{0.3cm}

\newpage
\noindent\textbf{I. Introduction}

\vspace{0.3cm}
  In the past few decades, the standard model (SM) of the particle physics has been evidenced to be a correct theory at the electroweak scale \cite{1}. It can successfully account for the vast majority of the particle and cosmological phenomena. However, the SM has also some shortcomings, namely, it can not explain some important issues such as the flavour puzzle \cite{2}, the tiny neutrino masses \cite{3}, the matter-antimatter asymmetry \cite{4}, the cold dark matter (CDM) \cite{5}. All kinds of theoretical ideas have been suggested to solve these problems all the time. The tiny neutrino masses can be implemented by the see-saw mechanism \cite{6}, or the radiative generation \cite{7}. The baryon asymmetry can be achieved by the electroweak baryogenesis \cite{8}, or the thermal leptogenesis \cite{9}. The CDM candidates are possibly the real scalar boson \cite{10}, the sterile neutrino \cite{11}, the lightest supersymmetric particle \cite{12}, the axion \cite{13}, and so on. Although many progresses on these fields have been made, a convincing and unified theory is not established as yet \cite{14}.

  The four things of the SM, the neutrino masses, the CDM and the matter-antimatter asymmetry appear to be not related to each other, this is hard to believe. In addition, it is well-known that there is a great desert between the GUT and the electroweak scale, this is very unnatural. Based on the universe harmony and the nature unification, there should be a transition theory between the both scales. It can not only accommodate the four things simultaneously, but also integrate them completely. On the other hand, the correct theory should keep such principles as the simplicity, the feasibility and the fewer number of parameters, moreover, it is promising to be tested in future experiments. If one theory is excessive complexity and unable to be tested, it is incredible and infeasible. For these purposes, I attempt to construct an effective model between the GUT and the electroweak scale. Its characteristics are a as small as possible  extension of the SM but able to integrate the four things. In any case, an investigation of new theory beyond the SM is always significant for particle physics as well as cosmology.

  The remainder of this paper is organized as follows. In Section II I outline the model. I will respectively discuss the matter-antimatter asymmetry and the cold dark matter in Sec. III and Sec. IV. The numerical results are given in Sec. V. Sec. VI is devoted to conclusions.

\vspace{1cm}
\noindent\textbf{II. Model}

\vspace{0.3cm}
 First of all, I assume the product groups $SU(3)_{C}\otimes SU(2)_{L}\otimes U(1)_{D}\otimes U(1)_{R}\otimes U(1)_{B-L}$ as the local gauge symmetries of the effective model between the GUT and the electroweak scale, where $D,R,B-L$ denote respectively the conserved quantum numbers of the three Abelian subgroups. Secondly, the model particle contents are divided into the active SM sector and the inactive Dark sector. Their notations and gauge quantum numbers are in detail listed by the following table,
\ba
\begin{array}{|c|l|l|}
 \hline  &\hspace{1.5cm}\mbox{SM sector} &\hspace{1.5cm}\mbox{Dark sector}\\
 \hline \mbox{gauge bosons} &\hspace{0.3cm}G_{\mu}^{a}\hspace{0.5cm}W_{\mu}^{i} &\hspace{0.3cm}X_{\mu}^{D}\hspace{0.5cm}X_{\mu}^{R}\hspace{0.5cm}X_{\mu}^{B-L}\\
 \hline \mbox{fermions and scalars} &\hspace{0.3cm}H\hspace{0.5cm}q_{L}\hspace{0.5cm}u_{R}\hspace{0.5cm}d_{R}\hspace{0.5cm}l_{L}\hspace{0.5cm}e_{R} &\hspace{0.3cm}\nu_{R}\hspace{0.5cm}\chi_{R}\hspace{0.4cm}\chi_{L}\hspace{0.4cm}\phi_{1}\hspace{0.5cm}\phi_{2}\hspace{0.5cm}\Phi\\
 \hline SU(3)_{C}\otimes SU(2)_{L} &(1,2)(3,2)(3,1)(3,1)(1,2)(1,1) &(1,1)(1,1)(1,1)(1,1)(1,1)(3,1)\\
 \hline U(1)_{D} &\hspace{0.4cm}0\hspace{0.7cm}0\hspace{0.7cm}0\hspace{0.7cm}0\hspace{0.7cm}0\hspace{0.7cm}0 &\hspace{0.4cm}1\hspace{0.2cm}-1\hspace{0.2cm}-2\hspace{0.2cm}-1\hspace{0.7cm}0\hspace{0.65cm}0\\
 \hline U(1)_{R} &\hspace{0.1cm}-1\hspace{0.65cm}0\hspace{0.7cm}1\hspace{0.2cm}-1\hspace{0.7cm}0\hspace{0.2cm}-1 &\hspace{0.4cm}0\hspace{0.7cm}1\hspace{0.7cm}2\hspace{0.75cm}1\hspace{0.7cm}1\hspace{0.15cm}-1\\
 \hline U(1)_{B-L} &\hspace{0.4cm}0\hspace{0.65cm}\frac{1}{3}\hspace{0.7cm}\frac{1}{3}\hspace{0.7cm}\frac{1}{3}\hspace{0.15cm}-1\hspace{0.2cm}-1 &\hspace{0.1cm}-1\hspace{0.65cm}0\hspace{0.7cm}0\hspace{0.75cm}0\hspace{0.2cm}-1\hspace{0.65cm}\frac{1}{3}\\
 \hline U(1)_{Y=D+R+(B-L)} &\hspace{0.1cm}-1\hspace{0.6cm}\frac{1}{3}\hspace{0.7cm}\frac{4}{3}\hspace{0.2cm}-\frac{2}{3}\hspace{0.2cm}-1\hspace{0.15cm}-2 &\hspace{0.4cm}0\hspace{0.7cm}0\hspace{0.7cm}0\hspace{0.75cm}0\hspace{0.7cm}0\hspace{0.15cm}-\frac{2}{3}\\
 \hline
\end{array}
\nonumber
\ea
where all kinds of the notations are self-explanatory, $U(1)_{Y}$ is the supercharge subgroup of the SM, it is derived from the model symmetry breakings. Obviously, the fermions and scalars in the SM sector have no charges of $U(1)_{D}$, in contrast, the ones in the Dark sector are all singlets under the SM groups except the only leptoquark $\Phi$ which is a coloured scalar. Note that the three $U(1)$ subgroups completely determinate three relative phase transformations among the four chiral fermions, $e_{R},\nu_{R},\chi_{R},\chi^{c}_{R}(\chi_{L})$. The right-handed neutrino $\nu_{R}$ is a Dirac nature lepton in the model. The other neutral Dirac fermion $\chi$ is different from the quarks and also different from the leptons, it has a vanishing $B-L$ number, in fact, it is namely the CDM in the model. For simplicity, $\chi$ is assumed as only one generation hereinafter.

  Thirdly, it can be seen by the above table that the model symmetries will be broken according to the following chain,
\begin{alignat}{1}
 &U(1)_{R}\otimes U(1)_{B-L} \xrightarrow{\langle\phi_{2}\rangle\sim 10^{12}\mathrm{GeV}} U(1)_{R+(B-L)}\,, \nonumber\\
 &U(1)_{D}\otimes U(1)_{R+(B-L)} \xrightarrow{\langle\phi_{1}\rangle\sim 10^{3}\mathrm{GeV}} U(1)_{Y=D+R+(B-L)}\,, \nonumber\\
 &SU(2)_{L}\otimes U(1)_{Y} \xrightarrow{\langle H\rangle\sim 10^{2}\mathrm{GeV}} U(1)_{em}\,.
\end{alignat}
This breaking chain will be implemented by the following scalar potentials (6). In addition, the arrangements in the table imply that $B-L$ is anomaly-free. On the one hand, the cancellation of the Adler-Bell-Jackiw anomaly for the $SU(2)_{L}$ factor is as usual because the particles in the Dark sector are all singlets under $SU(2)_{L}$. On the other hand, a simple calculation shows that $U(1)_{B-L}$ is anomaly-free. Although both $U(1)_{D}$ and $U(1)_{R}$ are anomaly, $U(1)_{D+R}$ is anomaly-free. Therefore, $U(1)_{Y}$ is eventually anomaly-free.

  Lastly, the full model Lagrangian can be written out on the basis of the gauge symmetries and the particle contents. The gauge kinetic energy terms are
\begin{alignat}{1}
 \mathscr{L}_{Gauge}=
 &\:\mathscr{L}_{pure\,gauge}+\sum\limits_{f_{L}}i\,\overline{f_{L}}\,\gamma^{\mu}D_{\mu}f_{L}
  +\sum\limits_{f_{R}}i\,\overline{f_{R}}\,\gamma^{\mu}D_{\mu}f_{R} \nonumber\\
 &+(D^{\mu}H)^{\dagger}D_{\mu}H+(D^{\mu}\phi_{1})^{\dagger}D_{\mu}\phi_{1}
  +(D^{\mu}\phi_{2})^{\dagger}D_{\mu}\phi_{2}+(D^{\mu}\Phi)^{\dagger}D_{\mu}\Phi,
\end{alignat}
where $f_{L,R}$ denote all kinds of the fermions in the table, and the covariant derivative $D_{\mu}$ is defined by
\ba
 D_{\mu}=\partial_{\mu}+i\left(g_{s}G_{\mu}^{a}\frac{\lambda^{a}}{2}+g_{w}W_{\mu}^{i}\frac{\tau^{i}}{2}+g_{D}X_{\mu}^{D}\frac{Q^{D}}{2}+g_{R}X_{\mu}^{R}\frac{Q^{R}}{2}+g_{B-L}X_{\mu}^{B-L}\frac{B-L}{2}\right).
\ea
In (3), $g_{s},g_{w},y_{D},g_{R},g_{B-L}$ are corresponding gauge coupling coefficients, $\lambda^{a}$ and $\tau^{i}$ are respectively the Gell-Mann and Pauli matrices, $Q^{D},Q^{R}$ and $B-L$ are respectively the charge operators of $U(1)_{D},U(1)_{R}$ and $U(1)_{B-L}$.

  The effective Yukawa couplings are
\begin{alignat}{1}
 \mathscr{L}_{Yukawa}=
 &\:\overline{q_{L}}Y_{u}u_{R}H+\overline{q_{L}}Y_{d}d_{R}i\tau_{2}H^{*}+\overline{l_{L}}Y_{e}e_{R}i\tau_{2}H^{*}+\phi_{1}\overline{\chi_{L}}Y_{\chi}\chi_{R}+\frac{\phi_{1}}{\Lambda}\overline{l_{L}}Y_{\nu}\nu_{R}H \nonumber\\
 &+\frac{\phi_{2}}{\Lambda}\epsilon_{\alpha\beta\gamma}\Phi_{\alpha}(\frac{1}{2}q_{L\beta}^{T}Y_{1}i\tau_{2}q_{L\gamma}+u_{R\beta}^{T}Y_{2}d_{R\gamma})+\frac{\phi_{2}^{*}}{\Lambda}\Phi^{\dagger}(l_{L}^{T}Y_{3}i\tau_{2}q_{L}+e_{R}^{T}Y_{4}u_{R}) \nonumber\\
 &+\frac{\phi_{1}^{*}}{\Lambda}\Phi^{\dagger}\chi_{R}^{T}Y_{5}d_{R}+\frac{1}{2\Lambda}d_{R}^{T}\Phi^{*}Y_{6}\Phi^{\dagger}d_{R}+h.c.\,,
\end{alignat}
where the charge conjugation matrix $C$ is omitted, which should be sandwiched between two spinor fields in the second and third line terms. $i\tau_{2}$ is inserted so as to satisfy the $SU(2)_{L}$ isospin symmetry. $\epsilon_{\alpha\beta\gamma}$ is a totally antisymmetric three-tensor for the color indices, it is used to guarantee the $SU(3)_{C}$ color symmetry. $Y_{u},Y_{d},\cdots,Y_{6}$ are the Yukawa coupling matrices, in addition, both $Y_{1}$ and $Y_{6}$ are symmetric structures on account of the spinor properties. The Yukawa matrices are generally complex, however, some complex phases can not be removed by the flavour basis choice and the redefined field phases, therefore the Yukawa sector certainly provides new $CP$-violating sources besides the SM one. Finally, all of the 5-dimensional couplings are suppressed by the GUT scale of $\Lambda\sim 10^{16}$ GeV. These terms may arise from the breakings of some GUT models, in particular, the last term in the first line of (4) is the Dirac coupling of the neutrinos, which will give rise to the neutrino masses after both $U(1)_{D}$ and electroweak breakings.

  After $H$ and $\phi_{1}$ developing the vacuum expectation values (see the following equation (7)), the first line terms of (4) generate the Dirac masses of all kinds of the fermions,
\ba
m_{u}=-\frac{v_{H}}{\sqrt{2}}Y_{u},\hspace{0.3cm} m_{d}=\frac{v_{H}}{\sqrt{2}}Y_{d},\hspace{0.3cm} m_{e}=\frac{v_{H}}{\sqrt{2}}Y_{e},\hspace{0.3cm} M_{\chi}=-\frac{v_{1}}{\sqrt{2}}Y_{\chi},\hspace{0.3cm} m_{\nu}=-\frac{v_{1}v_{H}}{2\Lambda}Y_{\nu}.
\ea
In contrast with the $v_{H}$ scale quarks and charged leptons and the $v_{1}$ scale CDM $\chi$, obviously, the neutrinos obtain only tiny masses due to the $\Lambda$ suppression. On the other hand, the second line terms of (4) violate the baryon or lepton number after $\phi_{2}$ developing the vacuum expectation value, which will lead to the matter-antimatter asymmetry.

  The scalar potentials are
\begin{alignat}{1}
 V_{Scalar}=&\:\lambda_{\Phi}(\Phi^{\dagger}\Phi)^{2}+\lambda_{H}\left(H^{\dagger}H-\frac{\lambda_{H}v_{H}^{2}+c_{4}v_{1}^{2}+c_{5}v_{2}^{2}}{2\lambda_{H}}\right)^{2} \nonumber\\
 &+\lambda_{1}\left(\phi_{1}^{\dagger}\phi_{1}-\frac{\lambda_{1}v_{1}^{2}+c_{4}v_{H}^{2}+c_{6}v_{2}^{2}}{2\lambda_{1}}\right)^{2}+\lambda_{2}\left(\phi_{2}^{\dagger}\phi_{2}-\frac{\lambda_{2}v_{2}^{2}+c_{5}v_{H}^{2}+c_{6}v_{1}^{2}}{2\lambda_{2}}\right)^{2} \nonumber\\
 &+2\Phi^{\dagger}\Phi(c_{1}H^{\dagger}H+c_{2}\phi_{1}^{\dagger}\phi_{1}+c_{3}\phi_{2}^{\dagger}\phi_{2})+2H^{\dagger}H(c_{4}\phi_{1}^{\dagger}\phi_{1}+c_{5}\phi_{2}^{\dagger}\phi_{2})+2c_{6}\phi_{1}^{\dagger}\phi_{1}\phi_{2}^{\dagger}\phi_{2}.
\end{alignat}
Furthermore, the vacuum configurations are directly obtained by discussing the $V_{Scalar}$ extreme as follows,
\ba
 \langle\Phi\rangle=0,\hspace{0.5cm} \langle H\rangle =\frac{v_{H}}{\sqrt{2}}\left(\begin{array}{c}1\\0\end{array}\right),\hspace{0.5cm} \langle\phi_{1}\rangle=\frac{v_{1}}{\sqrt{2}}\,,\hspace{0.5cm}
\langle\phi_{2}\rangle=\frac{v_{2}}{\sqrt{2}}\,.
\ea
These vacuum expectation values are assumed such hierarchy as $v_{H}\approx 246\:\mathrm{GeV}<v_{1}\approx(3\sim 4)\:\mathrm{TeV}\ll v_{2}\approx 10^{12}\:\mathrm{GeV}$. $v_{H}$ is fixed by the electroweak physics, $v_{1}$ will be determined by the tiny masses of the neutrinos and the relic abundance of the CDM $\chi$, $v_{2}$ should be very close to the reheating temperature of the universe \cite{15}. The vacuum spontaneous breakings give rise to the masses of the scalar bosons,
\begin{alignat}{1}
 &M_{Scalar}^{2}=\left(\begin{array}{cccc}2\lambda_{H}v_{H}^{2}&2c_{4}v_{H}v_{1}&2c_{5}v_{H}v_{2}&0\\
&2\lambda_{1}v_{1}^{2}&2c_{6}v_{1}v_{2}&0\\&&2\lambda_{2}v_{2}^{2}&0\\
&&&c_{1}v_{H}^{2}+c_{2}v_{1}^{2}+c_{3}v_{2}^{2}\end{array}\right), \nonumber\\
 &M_{H}\approx v_{H}\sqrt{\frac{2(\lambda_{H}\lambda_{1}\lambda_{2}+2c_{4}c_{5}c_{6}-\lambda_{H}c_{6}^{2}-\lambda_{1}c_{5}^{2}-\lambda_{2}c_{4}^{2})}{\lambda_{1}\lambda_{2}-c_{6}^{2}}}\,, \nonumber\\
 &M_{\phi_{1}}\approx v_{1}\sqrt{\frac{2(\lambda_{1}\lambda_{2}-c_{6}^{2})}{\lambda_{2}}}\,,\hspace{0.5cm} M_{\phi_{2}}\approx v_{2}\sqrt{2\lambda_{2}}\,,\hspace{0.5cm} M_{\Phi}\approx v_{2}\sqrt{c_{3}}\,, \nonumber\\
 &\left\vert\begin{array}{cccc}\lambda_{H}&c_{4}&c_{5}&0\\c_{4}&\lambda_{1}&c_{6}&0\\
c_{5}&c_{6}&\lambda_{2}&0\\0&0&0&c_{3}\\\end{array}\right\vert \:\mbox{is positive definite}.
\end{alignat}
The vacuum stability condition is namely that the determinant consisting of the coupling parameters is positive definite (it means that all of the ordered principal minors of the determinant are positive). The parameters are therefore restricted such that $[\lambda_{H},\lambda_{1},\lambda_{2},c_{3}]$ are all positive and $\sim\mathscr{O}(0.1)$, and $[c_{4},c_{5},c_{6}]$ are sufficiently small. However, we can always choose a set of suitable values to satisfy the conditions.

 Finally, the gauge symmetry breakings lead to the masses and mixings of the gauge bosons as follows,
\begin{alignat}{1}
 &g_{D}X_{\mu}^{D}\frac{Q^{D}}{2}+g_{R}X_{\mu}^{R}\frac{Q^{R}}{2}+g_{B-L}X_{\mu}^{B-L}\frac{B-L}{2}\longrightarrow \nonumber\\
 &g_{Y}B_{\mu}\frac{Y}{2}+\frac{g_{Y}}{sin\theta_{1}}X_{\mu}(cos\theta_{1}\frac{Y}{2}-\frac{1}{cos\theta_{1}}\frac{Q^{D}}{2})+\frac{g_{Y}}{sin\theta_{1}}X'_{\mu}(-tan\theta_{2}\frac{Q^{R}}{2}+\frac{1}{tan\theta_{2}}\frac{B-L}{2}), \nonumber\\
 &\left(\begin{array}{c}B_{\mu}\\X_{\mu}\\X'_{\mu}\end{array}\right)=\left(\begin{array}{ccc}cos\theta_{1}&sin\theta_{1}&0\\-sin\theta_{1}&cos\theta_{1}&0\\0&0&1\end{array}\right)\left(\begin{array}{ccc}1&0&0\\0&cos\theta_{2}&sin\theta_{2}\\0&-sin\theta_{2}&cos\theta_{2}\end{array}\right)\left(\begin{array}{c}X_{\mu}^{D}\\X_{\mu}^{R}\\X_{\mu}^{B-L}\end{array}\right), \nonumber\\
 &Y=Q^{D}+Q^{R}+(B-L),\hspace{0.3cm} g_{Y}^{-2}=g_{D}^{-2}+g_{R}^{-2}+g_{B-L}^{-2},\hspace{0.3cm} cos\theta_{1}=\frac{g_{Y}}{g_{D}}\,,\hspace{0.3cm} tan\theta_{2}=\frac{g_{R}}{g_{B-L}}\,, \nonumber\\
 &M_{B_{\mu}}=0,\hspace{0.5cm} M_{X_{\mu}}=\frac{v_{1}g_{Y}}{sin2\theta_{1}}\sqrt{1+cos^{4}\theta_{1}\frac{v_{H}^{2}}{v_{1}^{2}}}\,,\hspace{0.5cm} M_{X'_{\mu}}=\frac{v_{2}g_{Y}}{sin\theta_{1}sin2\theta_{2}}\,, \nonumber\\
 &M_{W_{\mu}}=\frac{g_{w}v_{H}}{2},\hspace{0.5cm} M_{Z_{\mu}}=\frac{g_{w}v_{H}}{2cos\theta_{W}}\sqrt{1-cos^{4}\theta_{1}\frac{v_{H}^{2}}{v_{1}^{2}}}\,,
\end{alignat}
where the new gauge fields $B_{\mu},X_{\mu},X'_{\mu}$ are three mass eigenstates, $\theta_{1}$ and $\theta_{2}$ are respectively the mixing angles associated with the breakings of $\phi_{1}$ and $\phi_{2}$, $\theta_{W}$ is the weak-mixing angle. Note that the mixing angle between $Z_{\mu}$ and $X_{\mu}$ is $\sim\frac{v_{H}^{2}}{v_{1}^{2}}\sim 10^{-2}$, which is too small and can be neglected.

\vspace{1cm}
\noindent\textbf{III. Matter-antimatter Asymmetry}

\vspace{0.3cm}
  The model can naturally account for the matter-antimatter asymmetry. At the reheating temperature of the universe, the $B-L$ symmetry breaking due to $\langle\phi_{2}\rangle$ gives rise to the $v_{2}$ scale masses of the neutral $\phi_{2}$ and coloured $\Phi$, on the other hand, this also leads that the second line terms in (4) violate one unit of the $B-L$ number. In the light of the Yukawa couplings, $\Phi$ has three decay modes, namely $\Phi\rightarrow\overline{u}+\overline{d}$, $\Phi\rightarrow u+e$ and $\Phi\rightarrow d+\nu$. The first decay violates ``$-1$" unit of the baryon number, the last two decays violate ``$+1$" unit of the lepton number. The Feynman diagrams of (a) $\Phi\rightarrow\overline{u}+\overline{d}$ and (b) $\Phi\rightarrow u+e$ are shown in Figure 1, respectively.
\begin{figure}
 \centering
 \includegraphics[totalheight=8cm]{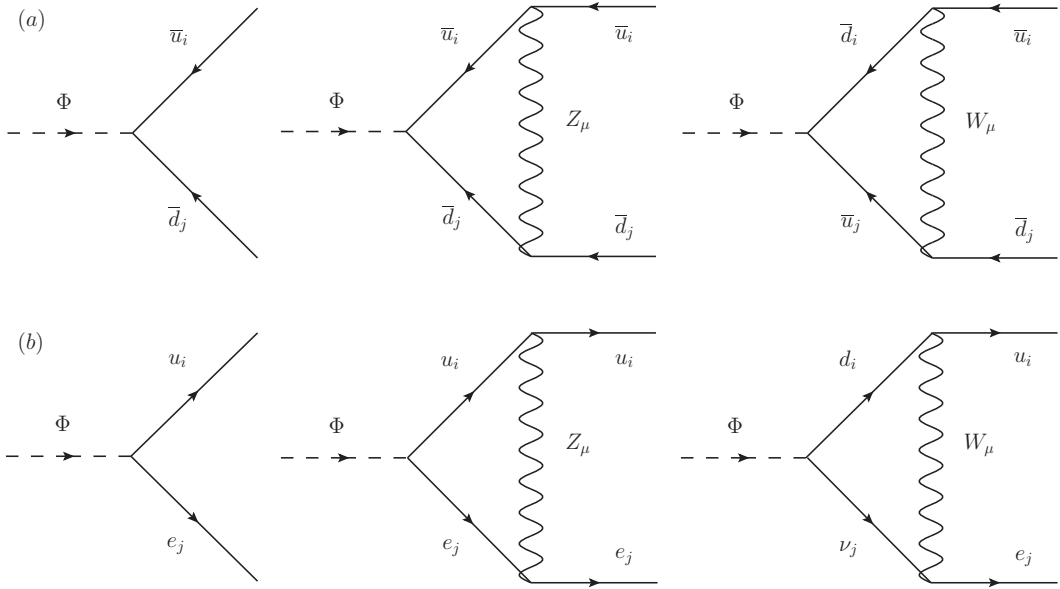}
 \caption{The tree and loop diagrams of (a) $\Phi\rightarrow\overline{u}+\overline{d}$ and (b) $\Phi\rightarrow u+e$, which lead to the matter-antimatter asymmetry.}
\end{figure}
One $CP$ asymmetry of each decay rate is generated by the interference between the tree diagram and the loop ones. The Feynman amplitude of $\Phi\rightarrow\overline{u}+\overline{d}$ is given by
\begin{alignat}{1}
 &\mathscr{M}=\frac{v_{2}}{\sqrt{2}\Lambda}v^{T}(p_{u})\left[(Y_{1}L+Y_{2}R)-\frac{g_{w}^{2}C_{loop}}{16\pi^{2}}(\frac{Q_{u_{L}}Q_{d_{L}}Y_{1}L+Q_{u_{R}}Q_{d_{R}}Y_{2}R}{M_{Z}^{2}}+\frac{Y_{1}L}{2M_{W}^{2}})\slashed{p}_{u}\slashed{p}_{d}\right]v(p_{d}), \nonumber\\
 &L,R=\frac{1\mp\gamma_{5}}{2},\hspace{0.5cm} Im[C_{loop}]=M_{\Phi}^{2}Im[c_{22}]+2Im[c_{24}]=Im[B_{0}(M_{\Phi}^{2},m_{u}^{2},m_{d}^{2})]=-i\pi, \nonumber\\
 &Q_{u_{L}}=\frac{1}{2cos\theta_{W}}-\frac{2sin^{2}\theta_{W}}{3cos\theta_{W}}\,,\hspace{0.5cm} Q_{u_{R}}=-\frac{2sin^{2}\theta_{W}}{3cos\theta_{W}}\,,\nonumber\\
 &Q_{d_{L}}=-\frac{1}{2cos\theta_{W}}+\frac{sin^{2}\theta_{W}}{3cos\theta_{W}}\,,\hspace{0.5cm} Q_{d_{R}}=\frac{sin^{2}\theta_{W}}{3cos\theta_{W}}\,,
\end{alignat}
where only the imaginary part of the loop integration factor $C_{loop}$ is relevant to the following $CP$ asymmetry. One can write out the parallel formulas for $\Phi\rightarrow u+e$.

  The decay $CP$-asymmetries of (a) and (b) in Figure 1 are defined and calculated as follows, respectively,
\begin{alignat}{1}
 &\varepsilon_{a}=\frac{\Gamma(\Phi\rightarrow\overline{u}+\overline{d})-\Gamma(\Phi^{*}\rightarrow u+d)}{\Gamma(\Phi)}=\frac{(\frac{1}{2}+sin^{2}\theta_{W})\sum\limits_{i,j}Im[(Y_{1})_{ij}(Y_{2}^{*})_{ij}]\frac{m_{ui}m_{dj}}{v_{H}^{2}}}{\pi Tr[2Y_{1}Y_{1}^{\dagger}+2Y_{2}Y_{2}^{\dagger}+2Y_{3}Y_{3}^{\dagger}+Y_{4}Y_{4}^{\dagger}]}\,,\nonumber\\
 &\varepsilon_{b}=\frac{\Gamma(\Phi\rightarrow u+e)-\Gamma(\Phi^{*}\rightarrow \overline{u}+\overline{e})}{\Gamma(\Phi)}=\frac{(\frac{3}{4}-\frac{5}{6}sin^{2}\theta_{W})\sum\limits_{i,j}Im[(Y_{3}^{\dagger})_{ij}(Y_{4}^{T})_{ij}]\frac{m_{ui}m_{ej}}{v_{H}^{2}}}{\pi Tr[2Y_{1}Y_{1}^{\dagger}+2Y_{2}Y_{2}^{\dagger}+2Y_{3}Y_{3}^{\dagger}+Y_{4}Y_{4}^{\dagger}]}\,,\nonumber\\
 &\Gamma(\Phi)=\Gamma(\Phi\rightarrow\overline{u}+\overline{d})+\Gamma(\Phi\rightarrow u+e)+\Gamma(\Phi\rightarrow d+\nu),
\end{alignat}
where $i,j$ are the generation indices of the fermions. The $CP$ asymmetries of (11) have the following characteristics. (i) $\varepsilon_{a}$ arises from the interference between $\Phi\rightarrow\overline{u}_{L}+\overline{d}_{L}$ and $\Phi\rightarrow\overline{u}_{R}+\overline{d}_{R}$, similarly, $\varepsilon_{b}$ results from the interference between $\Phi\rightarrow u_{L}+e_{L}$ and $\Phi\rightarrow u_{R}+e_{R}$, so (11) refers to the product factors $(Y_{1})_{ij}(Y_{2}^{*})_{ij}$ and $(Y^{\dagger}_{3})_{ij}(Y^{T}_{4})_{ij}$. However, the decay $\Phi\rightarrow d+\nu$ can not lead to a similar asymmetry because $\Phi\rightarrow d_{R}+\nu_{R}$ is non-existent. (ii) Under the mass eigenstate basis of the fermions, some irremovable complex phases in $Y_{1},Y_{2},Y_{3},Y_{4}$ are new $CP$-violating sources, they lead to $\varepsilon_{a}$ and $\varepsilon_{b}$ non-vanishing. (iii) The vertexes of the weak gauge bosons and fermions in Figure 1 contain the axil vector current couplings, in other words, the left-handed current coupling and the right-handed one are different sizes, or else the asymmetries will be vanishing. Obviously, only $Z_{\mu}^{0},W_{\mu}^{\pm}$ satisfy this condition, none of $X_{\mu},photon,gluon$ satisfies this. (iv) The sizes of $\varepsilon_{a}$ and $\varepsilon_{b}$ mainly depend on the ratio of the mass product of two final state fermions and the weak gauge boson squared masses (which are replaced by $v_{H}^{2}$ in (11)), moreover, they have no relation with $\Lambda,v_{2},M_{\Phi}$. On account of the mass hierarchy of the quarks and charged leptons, thus the factors such as $\frac{m_{u}m_{b}}{v_{H}^{2}},\frac{m_{u}m_{\tau}}{v_{H}^{2}}$ in (11) can naturally give rise to $(\varepsilon_{a},\varepsilon_{b})\sim 10^{-8}$, which eventually determine the size of the matter-antimatter asymmetry. It follows that there is a close relationship between the matter-antimatter asymmetry and the flavour physics.

  A simply estimate shows that the decay rates of $\Phi\rightarrow\overline{u}+\overline{d}$ and $\Phi\rightarrow u+e$ are far smaller than the Hubble expansion rate of the universe, namely
\ba
\Gamma(\Phi\rightarrow \overline{u}+\overline{d})=\frac{M_{\Phi}v_{2}^{2}}{16\pi\Lambda^{2}}Tr[Y_{1}Y_{1}^{\dagger}+Y_{2}Y_{2}^{\dagger}]\ll H(M_{\Phi})=\frac{1.66\sqrt{g_{*}}M_{\Phi}^{2}}{M_{Pl}}\,,
\ea
where $M_{Pl}=1.22\times10^{19}$ GeV. At the temperature of $M_{\Phi}$, all of the model particles are relativistic except $\Phi$ and $\phi_{2}$, so the effective number of relativistic degrees of freedom is $g_{*}=119.5$ in (12). Provided $M_{\Phi}\approx v_{2}$ and $Tr[Y_{1}Y_{1}^{\dagger}]\approx Tr[Y_{2}Y_{2}^{\dagger}]\approx 1$, one can estimate $\frac{\Gamma(\Phi)}{H}\lesssim\frac{v_{2}M_{pl}}{10^{2}\Lambda^{2}}\lesssim 10^{-3}$, therefore the decays are severely out-of-equilibrium processes. The above discussions are collected together, Sakharov's three conditions are completely satisfied \cite{16}, consequently, the two decays in Figure 1 can indeed generate the $B-L$ asymmetry.

  Because the $B-L$ asymmetry arises above the electroweak scale, the sphaleron process can efficiently convert it into the baryon asymmetry \cite{17}. The related expressions are
\ba
Y_{B}=\frac{n_{B}-\overline{n}_{B}}{s}=c_{s}Y_{B-L}=c_{s}\frac{(-1)(\varepsilon_{a}+\varepsilon_{b})}{g_{*}}\,,\hspace{0.3cm} \eta_{B}=7.04Y_{B}\approx 6.15\times10^{-10}\,,
\ea
where $c_{s}=\frac{28}{79}$ and $g_{*}=119.5$, and $7.04$ is a ratio of the entropy density to the photon number density. Note that only the SM particles are involved in the sphaleron process. In addition, the dilution effect can completely be ignored because the decays are seriously departure from thermal equilibrium. At the present day the experimental value of the baryon asymmetry, $\eta_{B}\sim6.15\times10^{-10}$, has been established by multiple approaches \cite{18}. The charged lepton asymmetry is equal to the proton one because of the electric neutrality of the universe. Only the neutrino asymmetry has been unknown so far.

\vspace{1cm}
\noindent\textbf{IV. Cold Dark Matter}

\vspace{0.3cm}
  At the TeV scale, the $U(1)_{D}$ symmetry breaking due to $\langle\phi_{1}\rangle$ gives rise to the $v_{1}$ scale masses of $\chi,\phi_{1},X_{\mu}$. The three new particles are all singlets under the SM groups, but they play key roles in the new physics beyond the SM. $\chi$ is a neutral Dirac fermion. It has not any direct couplings to the SM particles, but it can indirectly connect with them via the two mediators of $X_{\mu}$ and $\phi_{1}$. In addition, $\chi$ is a stable particle. Its stability is an inevitable outcome of the model gauge symmetries, we need not an extra symmetry to guarantee it. Therefore, $\chi$ is exactly a typical WIMP, moreover, it is a desirable candidate of the CDM.

  In the model, the current relic abundance of $\chi$ can be calculated by the thermal production in the early universe. In the light of the model Lagrangian and the results of the symmetry breakings, a pair of $\chi$ can annihilate into the SM particles via the $s$-channel mediation of $X_{\mu}$ or $\phi_{1}$, shown as (a) and (b) in Figure 2.
\begin{figure}
 \centering
 \includegraphics[totalheight=4.5cm]{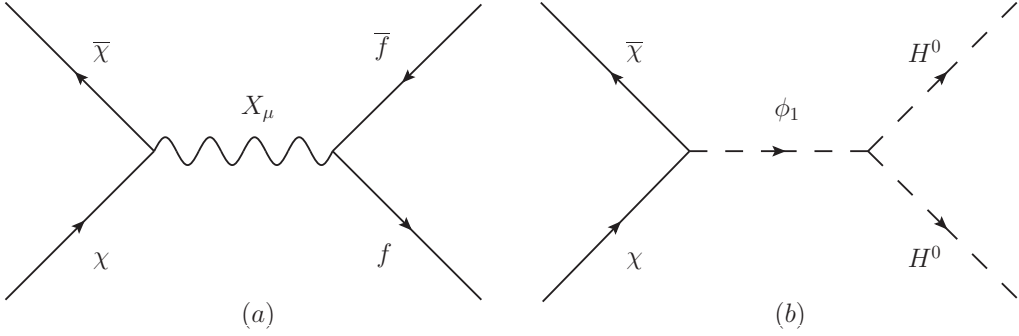}
 \caption{A pair of the CDM $\chi$ annihilating into the SM particles via the $s$-channel mediation of (a) $X_{\mu}$ or (b) $\phi_{1}$.}
\end{figure}
The cross-sections of (a) and (b) in Figure 2 are calculated as follows,
\begin{alignat}{1}
 &\sigma_{a}v=\frac{A(\frac{g_{Y}}{sin\theta_{1}})^{4}(s+B M_{\chi}^{2})}{384\pi(s-M_{X_{\mu}}^{2})^{2}}=\frac{Acos^{4}\theta_{1}M_{\chi}^{2}}{24\pi v_{1}^{4}}[\frac{4+B}{(1-4y)^{2}}+\frac{4+B-(2+B)(1-4y)}{2(1-4y)^{3}}\,v^{2}+\cdots], \nonumber\\
 &\sigma_{b}v=\frac{c_{4}^{2}M_{\chi}^{2}}{16\pi(s-M_{\phi_{1}}^{2})^{2}}(1-\frac{2M_{\chi}^{2}}{s})=\frac{c_{4}^{2}M_{\chi}^{2}}{32\pi M_{\phi_{1}}^{4}}[\frac{1}{(1-4y')^{2}}+\frac{1+4y'}{4(1-4y')^{3}}\,v^{2}+\cdots], \nonumber\\
 &A=(D^{2}_{\chi_{L}}+D^{2}_{\chi_{R}})\sum\limits_{f}(Y_{f_{L}}^{2}+Y_{f_{R}}^{2}),\hspace{0.3cm} B=2-3\frac{(D_{\chi_{L}}-D_{\chi_{R}})^{2}}{D^{2}_{\chi_{L}}+D^{2}_{\chi_{R}}},\hspace{0.3cm} y=\frac{M_{\chi}^{2}}{M_{X_{\mu}}^{2}}\,,\hspace{0.3cm} y'=\frac{M_{\chi}^{2}}{M_{\phi_{1}}^{2}}\,,
\end{alignat}
where $D_{\chi_{L,R}}$ are the $D$ number of $\chi_{L,R}$ and $Y_{f_{L,R}}$ are the supercharge number of the SM fermion, see the table. $v=2\sqrt{1-\frac{4M_{\chi}^{2}}{s}}$ is a relative velocity of two annihilating particles and s is the squared center-of-mass energy. Since $M_{\chi}$ and $M_{\phi_{1}}$ being proportional to $v_{1}$, essentially, the two cross-sections are inversely proportional to $v_{1}^{2}$, so one can roughly estimate $(\sigma_{a}v,\sigma_{b}v)\sim 10^{-9}\:\mathrm{GeV}^{-2}$ provided that the total contribution of the related parameters is $\sim 0.1$. The freeze-out temperature of $\chi$ is solved by the following equations,
\begin{alignat}{1}
 &\langle(\sigma_{a}+\sigma_{b})v\rangle_{T_{f}}n_{\chi}(T_{f})=H(T_{f})=\frac{1.66\sqrt{g_{*}(T_{f})}\,T_{f}^{2}}{M_{Pl}}\,,\nonumber\\
 &\langle(\sigma_{a}+\sigma_{b})v\rangle_{T_{f}}\approx a+b\,\langle v^{2}\rangle=a+6b\,\frac{T_{f}}{M_{\chi}}\,,\hspace{0.3cm} n_{\chi}(T_{f})=4\left( \frac{M_{\chi}T_{f}}{2\pi}\right)^{\frac{3}{2}}e^{-\frac{M_{\chi}}{T_{f}}},
\end{alignat}
where $a$ and $b$ are obtained by the expansion coefficients in (14). Finally, the relic abundance of $\chi$ is determined by the so-called ``WIMP Miracle" as follows \cite{19},
\ba
\Omega_{\chi}h^{2}=\frac{0.85\times10^{-10}\,\mathrm{GeV}^{-2}}{\sqrt{g_{*}(T_{f})}\,x(a+3b\,x)}\approx 0.12,\hspace{0.5cm} x=\frac{T_{f}}{M_{\chi}}\approx\frac{1}{17+lnM_{\chi}-\frac{3}{2}lnx}\approx\frac{1}{30}\,,
\ea
where $g_{*}(T_{f})=91.5$. Provided $M_{\chi}\sim 300$ GeV (which is determined by the following numerical calculation), then $T_{f}$ is $\sim 10$ GeV, at this temperature the relativistic particles include all of the particles whose masses are below $M_{W}$, thus one can figure out $g_{*}(T_{f})=91.5$. In short, we can correctly fit the current abundance of the CDM as long as the parameters are a set of suitable values.

  Now I show the stability of the CDM $\chi$ and proton by some simple calculations. In fact, the couplings in (4) can lead to the decays of $\chi$ and proton via the mediation of the leptoquark $\Phi$, namely there are processes such as $\chi\rightarrow\overline{n}^{0},\chi\rightarrow\pi^{+}+e^{-},\chi\rightarrow\pi^{0}+\nu^{0}$ and $p^{+}\rightarrow\pi^{0}+e^{+},p^{+}\rightarrow\pi^{+}+\overline{\nu}^{0}$, shown as (a) and (b) in Figure 3.
\begin{figure}
 \centering
 \includegraphics[totalheight=4.5cm]{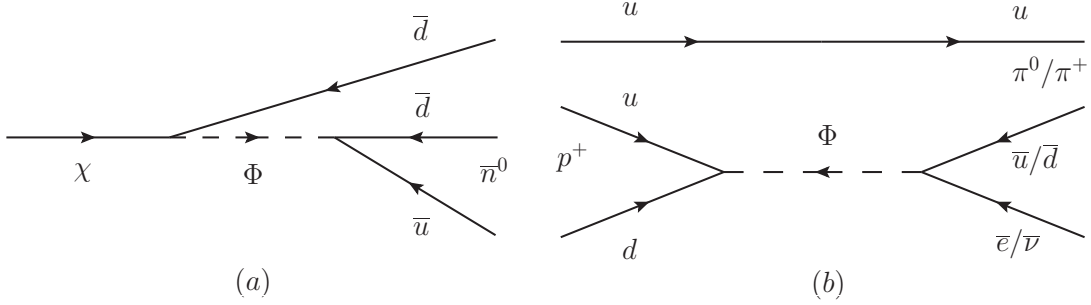}
 \caption{(a) The CDM $\chi$ decay into one anti-neutron, (b) the proton decay into $\pi^{0}+e^{+}$ or $\pi^{+}+\overline{\nu}^{0}$.}
\end{figure}
The decay widths of (a) and (b) in Figure 3 are given by
\begin{alignat}{1}
 &\Gamma(\chi)=\frac{3M_{\chi}^{5}v_{1}^{2}v_{2}^{2}}{128^{2}\pi^{3} M_{\Phi}^{4}\Lambda^{4}}\vert (Y_{5})_{1}\vert^{2}\left(\vert(Y_{1})_{11}\vert^{2}+\vert(Y_{2})_{11}\vert^{2}\right), \nonumber\\
 &\Gamma(p)=\frac{m_{p}^{5}v_{2}^{4}}{256\pi M_{\Phi}^{4}\Lambda^{4}}\left(\vert(Y_{1})_{11}\vert^{2}+\vert(Y_{2})_{11}\vert^{2}\right)\left(\vert(Y_{3})_{11}\vert^{2}+\vert(Y_{4})_{11}\vert^{2}\right), \nonumber\\
 &\frac{\tau(\chi)}{\tau(p)}\sim 10^{2}\frac{m_{p}^{5}v_{2}^{2}}{M_{\chi}^{5}v_{1}^{2}}\gg 1.
\end{alignat}
Provided $M_{\Phi}\approx v_{2}$ and $\vert(Y_{i})_{11}\vert^{2}\lesssim 0.1$, then the proton lifetime is $\tau(p)\gtrsim 10^{36}$ year, in addition, the $\chi$ lifetime is far larger than the proton one provided $v_{2}\sim 10^{12}$ GeV and $v_{1}\sim 10^{3}$ GeV. These results are very well in accordance with the current experimental limit \cite{20}. In a word, both the CDM $\chi$ and proton are very stable in the model.

\vspace{1cm}
\noindent\textbf{V. Numerical Results and Discussions}

\vspace{0.3cm}
  In the section I present the numerical results of the model. The fundamental parameters of the model are chosen as the typical values as follows,
\begin{alignat}{1}
 &\Lambda=10^{16}\:\mathrm{GeV},\hspace{0.5cm} v_{2}=10^{12}\:\mathrm{GeV},\hspace{0.5cm} v_{1}=3.5\:\mathrm{TeV},\hspace{0.5cm} v_{H}=246\:\mathrm{GeV}, \nonumber\\
 &g_{Y}=0.356,\hspace{0.3cm} sin\theta_{1}=0.3,\hspace{0.3cm} Y_{\chi}=0.124,\hspace{0.3cm} Y_{\nu}=1,\hspace{0.3cm} c_{4}=0.1,\hspace{0.3cm} M_{\phi_{1}}=750\:\mathrm{GeV},
\end{alignat}
where I only consider one generation of $\chi$ and $\nu$ for the sake of simplicity. $v_{H}$ and $g_{Y}$ are fixed by the SM data. The values of $\Lambda$ and $v_{2}$ are required by the model. The current experimental data indicate that the masses of the neutrinos are only $\sim 0.01$ eV \cite{20}, in addition, $Y_{\nu}\approx 1$ is a reasonable value, so one can infer $v_{1}\approx(3\sim 4)$ TeV by use of (5). $c_{4}\approx 0.1$ is also very natural for the scalar couplings. According to (8), $M_{\phi_{1}}$ should be smaller than $v_{1}$ and has a greater parameter space. Here I suppose that $\phi_{1}$ is likely the new boson detected recently at the LHC (the reason for this will be discussed later), so I take $M_{\phi_{1}}=750$ GeV. Provided $g_{D}<1$ and $M_{X_{\mu}}<v_{1}$, by use of (9) one can obtain the parameter areas of $0.18\lesssim sin\theta_{1}\lesssim 0.94$. The value of $M_{X_{\mu}}$ varies with $sin\theta_{1}$. Finally, the value of $Y_{\chi}$, which directly determines $M_{\chi}$, is given by fitting the observed value $\Omega_{\chi}h^{2}\approx 0.12$. Now (18) is input into (5), (9) and (14)-(16), we thus obtain the following results,
\ba
M_{X_{\mu}}=2.18\:\mathrm{TeV},\hspace{0.5cm} M_{\chi}=308\:\mathrm{GeV},\hspace{0.5cm} m_{\nu}=0.043\:\mathrm{eV},\hspace{0.5cm} \Omega_{\chi}h^{2}=0.12.
\ea
These are very well in agreement with the current experimental data \cite{20}.

  Figure 4 shows that both $M_{X_{\mu}}$ and $M_{\chi}$ are subject to $sin\theta_{1}$ for $v_{1}=4,\,3.5,\,3$ TeV, respectively.
\begin{figure}
 \centering
 \includegraphics[totalheight=8cm]{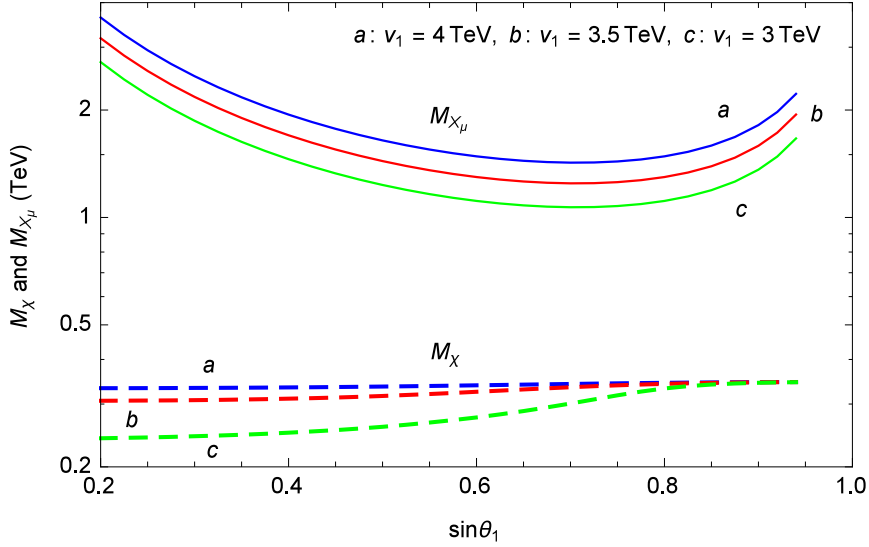}
 \caption{The graph of $M_{X_{\mu}}$ and $M_{\chi}$ being subject to $sin\theta_{1}$, (a) blue curves, (b) red curves and (c) green curves correspond with $v_{1}=4,\,3.5,\,3$ TeV, respectively, the rest of the parameters are fixed as (18).}
\end{figure}
It is clearly seen from the graph that $M_{X_{\mu}}$ has a lot of uncertainty in the area of $1\:\mathrm{TeV}\lesssim M_{X_{\mu}}\lesssim 3.5\:\mathrm{TeV}$, but the CDM $\chi$ mass is only in the narrow area of $250\:\mathrm{GeV}\lesssim M_{\chi}\lesssim 350\:\mathrm{GeV}$. When $sin\theta_{1}$ approaches to the left minimal value, $\Omega_{\chi}h^{2}$ is mostly dominated by $\sigma_{a}$, whereas when $sin\theta_{1}$ is close to the right maximal value, $\Omega_{\chi}h^{2}$ is mostly dominated by $\sigma_{b}$. In overall, the most reasonable value of $sin\theta_{1}$ is probably around $sin\theta_{1}\approx 0.3$. In brief, $X_{\mu}$ and $\chi$ are the two new particles beyond the SM, they are expected to be discovered in the future experiments.

  In order to calculate the baryon asymmetry, we need chose values of the Yukawa couplings $Y_{1,2,3,4}$ and input the masses of the quarks and charged leptons in terms of (11). The detailed values are taken as follows (in GeV as mass unit) \cite{20},
\begin{alignat}{1}
 &m_{u}=0.0023,\hspace{0.5cm} m_{c}=1.275,\hspace{0.5cm} m_{t}=173, \nonumber\\
 &m_{d}=0.0048,\hspace{0.5cm} m_{s}=0.095,\hspace{0.5cm} m_{b}=4.18, \nonumber\\
 &m_{e}=0.000511,\hspace{0.5cm} m_{\mu}=0.1057,\hspace{0.5cm} m_{\tau}=1.777, \nonumber\\
 &sin^{2}\theta_{W}=0.231,\hspace{0.5cm} Tr[Y_{i}Y_{i}^{\dagger}]=1, \nonumber\\
 &Im[(Y_{1})_{22}(Y_{2}^{*})_{22}]=-0.44,\hspace{0.5cm} Im[(Y_{1})_{31}(Y_{2}^{*})_{31}]=-0.065, \nonumber\\
 &Im[(Y_{3}^{\dagger})_{22}(Y_{4}^{T})_{22}]=-0.52,\hspace{0.5cm} Im[(Y_{3}^{\dagger})_{31}(Y_{4}^{T})_{31}]=-0.8,
\end{alignat}
where $i=1,2,3,4$. Under the mass eigenstate basis of the fermions, the imaginary parts of $Y_{1,2,3,4}$ are not all zero. In view of the mass hierarchy of the quarks and charged leptons, a careful analysis shows that the four decays, $\Phi\rightarrow\overline{c}+\overline{s}$, $\Phi\rightarrow\overline{t}+\overline{d}$, $\Phi\rightarrow c+\mu$, $\Phi\rightarrow t+e$, are preferred because any one of them can separately fit $\eta_{B}\approx 6.15\times10^{-10}$. The four sets of Yukawa coupling values in (20) respectively correspond with the four decay modes. Of course, the baryon asymmetry is possibly attributed to the total contribution of them.

  To sum up, all of the numerical results, which are naturally produced without any fine tuning, can completely fit all kinds of the experimental data. This clearly demonstrate that this model is reasonable and feasible.

  In the end, I simply discuss the test of the model. The three new particles of the model, $\chi,\phi_{1},X_{\mu}$, are able to be produced at the TeV-scale colliders. The specific processes are
\begin{alignat}{1}
 &e^{-}+e^{+}\rightarrow X_{\mu}\rightarrow\chi+\overline{\chi},\hspace{0.5cm} p+\overline{p}\rightarrow X_{\mu}\rightarrow\chi+\overline{\chi}, \nonumber\\
 &p+p\rightarrow X_{\mu}+X_{\mu}\rightarrow\phi_{1},\hspace{0.5cm} \phi_{1}\rightarrow \chi+\overline{\chi}\:\mathrm{or}\:H+H.
\end{alignat}
Both $e^{-}+e^{+}$ and $p+\overline{p}$ can produce a pair of the CDM $\chi$ via the s-channel mediation of $X_{\mu}$. Although their cross-sections are only $\sim 10^{-9}\:\mathrm{GeV}^{-2}$ due to the heavy $M_{X_{\mu}}$, we are very promising to find $\chi$ and $X_{\mu}$ in the near future. At present we have an opportunity to find $\phi_{1}$ via two $X_{\mu}$ fusion at the LHC \cite{21}, which can decay into a pair of the CDM $\chi$ or Higgs bosons. Recently, the $750$ GeV boson detected at the LHC is likely to be $\phi_{1}$. In a word, it is feasible to test the model in future collider experiments.

\vspace{1cm}
\noindent\textbf{VI. Conclusions}

\vspace{0.3cm}
  In summary, I suggest an effective theory between the GUT and the electroweak scale, which is a natural and reasonable extension of the SM. The new model introduces the two symmetries of $U(1)_{B-L}$ and $U(1)_{D}$, which are respectively broken at the reheating temperature of the universe and the TeV scale. The model particles consist of the SM ones and the dark sector. The particles in the dark sector include the right-handed Dirac neutrino, the CDM fermion, etc, they are all singlets under the SM groups except the super-heavy leptoquark $\Phi$, furthermore, they are directly relevant to the new physics beyond the SM. The model can clearly account for the origins of the tiny neutrino masses, the matter-antimatter asymmetry and the cold dark matter. In addition, the model gives some interesting results and predictions, for example, the neutrinos are Dirac nature and their masses are related to the $U(1)_{D}$ breaking, the size of the matter-antimatter asymmetry is closely related to the mass hierarchy of the quarks and charged leptons, the CDM $\chi$ mass is probably in the range of $250\sim 350$ GeV. Finally, the model is simple and feasible, we are promising to test it in future collider experiments.

\vspace{1cm}
 \noindent\textbf{Acknowledgments}

\vspace{0.3cm}
  I would like to thank my wife for her large helps. This research is supported by the Fundamental Research Funds for the Central Universities Grant No. WK2030040054.


\vspace{1cm}
\begin{thebibliography}{99}
\bibitem{1}
 G. Altarelli, M. W. Grunewald, Phys. Reps. 403-404, 189 (2004);
 C. Quigg, Annu. Rev. Nucl. Part. Sci. 59, 505 (2009).
\bibitem{2}
 Zhi-Zhong Xing, Int. J. Mod. Phys. A 29, 1430067 (2014);
 Z. Ligeti, arXiv:1502.01372.
\bibitem{3}
 R. N. Mohapatra, \emph{et al.}, Rep. Prog. Phys. 70, 1757 (2007);
 G. Altarelli, Int. J. Mod. Phys. A 29, 1444002 (2014).
\bibitem{4}
 M. Dine and A. Kusenko, Rev. Mod. Phys. 76, 1 (2004);
 L. Canetti, M. Drewes and M. Shaposhnikov, New J. Phys. 14, 095012 (2012).
\bibitem{5}
 V. Lukovic, P. Cabella and N. Vittorio, Int. J. Mod. Phys. A 29, 1443001 (2014);
 G. Bertone, Particle Dark Matter (Cambridge University Press, 2010).
\bibitem{6}
M. Gell-Mann, P. Ramond, R. Slansky, in Supergravity, eds. P. van Niewenhuizen and D. Z. Freeman (North-Holland, Amsterdam, 1979);
T. Yanagida, in Proc. of the Workshop on Unified Theory and Baryon Number in the Universe, eds. O. Sawada and A. Sugamoto (Tsukuba, Japan, 1979);
R. N. Mohapatra, G. Senjanovic, Phys. Rev. Lett. 44, 912 (1980).
\bibitem{7}
Andr´e de Gouvˆea, Annu. Rev. Nucl. Part. Sci. 66, 197 (2016);
Ernest Ma, Phys. Rev. Lett. 112, 091801 (2014).
\bibitem{8}
James M. Cline, arXiv:hep-ph/0609145;
D. E. Morrissey and M. J. Ramsey-Musolf, New J. Phys. 14, 125003 (2012).
\bibitem{9}
M. Fukugita, T. Yanagida, Phys. Lett. B 174, 45 (1986);
W. Buchmuller, R. D. Peccei and T. Yanagida, Annu. Rev. Nucl. Part. Sci. 55, 311 (2005);
S. Davidson, E. Nardi and Y. Nir, Phys. Reps. 466, 105 (2008).
\bibitem{10}
T. Bringmann, New J. Phys. 11, 105027 (2009);
C. P. Burgess, M. Pospelov and T. ter Veldhuis, Nucl.Phys.B619, 709 (2001).
\bibitem{11}
L. Canetti, M. Drewes and M. Shaposhnikov, Phys. Rev, Lett 110, 061801 (2003);
A. Boyarsky, O. Ruchayskiy, and M. Shaposhnikov, Annu. Rev. Nucl. Part. Sci. 59, 191 (2009);
A. Kusenko, Phys. Reps. 481, 1 (2009).
\bibitem{12}
G. Jungman, M. Kamionkowski, and K. Griest, Phys. Reps. 267, 195 (1996).
\bibitem{13}
David J.E. Marsh, Phys. Reps. 643, 1 (2016).
\bibitem{14}
 H. Davoudiasl and R. N. Mohapatra, New J. Phys. 14, 095011 (2012);
 J. Harz and Wei-Chi Huang, Int. J. Mod. Phys. A 30, 1530045 (2015);
 W. M. Yang, Nucl. Phys. B 885, 505 (2014);
 W. M. Yang, Phys. Rev. D 87, 095003 (2013).
\bibitem{15}
 R. Allahverdi, R. Brandenberger, F. Cyr-Racine, and A. Mazumdar, Annu. Rev. Nucl. Part. Sci. 60, 27 (2010).
\bibitem{16}
 A. D. Sakharov, Pisma Zh. Eksp. Teor. Fiz. 5, 32 (1967), JETP Lett. 5, 24 (1967), Sov. Phys.-Usp. 34, 392 (1991), Usp. Fiz. Nauk 161, 61 (1991).
\bibitem{17}
 V. A. Kuzmin, V. A. Rubakov, M. A. Shaposhnikov, Phys. Lett. B 155, 36 (1985);
 W. Buchmuller, R. D. Peccei and T. Yanagida, Annu. Rev. Nucl. Part. Sci. 55, 311 (2005)
\bibitem{18}
 E. Komatsu \emph{et al} [WMAP Collaboration], Astrophys. J. Suppl. 192, 18 (2011).
\bibitem{19}
 G. B. Gelmini, arXiv:1502.01320;
 D. Hooper, arXiv:0910.4090.
\bibitem{20}
 K. A. Olive \emph{et al.} [Particle Data Group], Chin. Phys. C, 38, 090001 (2014).
\bibitem{21}
 D. E. Morrissey, T. Plehn, T. M.P. Tait, Phys. Reps. 515, 1 (2012).
\end{thebibliography}
\end{document}